\documentclass[aps,preprint]{revtex4}
\usepackage{amsmath}
\usepackage{amssymb}
\usepackage{mathrsfs}
\usepackage{xcolor}
\usepackage{graphicx}
\usepackage{subfigure}
\usepackage{enumerate}
\usepackage{float}

\begin{document}
	%\preprint{CTP-SCU/2021010}	
	\title{Echoes from charged black holes influenced by quintessence}

	\author{Siyuan Hui$^{a,b}$}
	\email{huisiyuan@stu.scu.edu.cn}
	\author{Benrong Mu$^{a,b}$}
	\email{benrongmu@cdutcm.edu.cn}
	
	\affiliation{$^{a}$Center for Joint Quantum Studies, College of Medical Technology, Chengdu University of Traditional Chinese Medicine, Chengdu, 611137, PR China\\
	$^{b}$Center for Theoretical Physics, College of Physics, Sichuan University, Chengdu, 610064, PR China}

	\begin{abstract}
	
	In this paper, we investigate the effective potential and echoes from the dyonic black hole with quintessence. For a dyonic black hole, the quasi-topological electromagnetism provides proper matter energy-momentum tensor to curve the spacetime, and quintessence strengthens this force. We find that when the effect of quintessence becomes stronger, the black hole potential transforms between single-peak and double-peak, which will influence the existence of black hole echoes. In particular, we find that observer will receive a sudden vanishment of high-frequency echoes when quintessence remains a relatively strong effect.
	
	\end{abstract}
	\maketitle
	\tableofcontents
	
	\section{Introduction}

	LIGO and Virgo first successfully detected gravitational waves from a binary black hole merger\cite{LIGOScientific:2016aoc}. Later, the Event Horizon Telescope photographed the first image of a supermassive black hole at the center of galaxy M87\cite{EventHorizonTelescope:2019dse, EventHorizonTelescope:2019uob, EventHorizonTelescope:2019jan, EventHorizonTelescope:2019ths, EventHorizonTelescope:2019pgp, EventHorizonTelescope:2019ggy}. This discovery launches an extraordinary new area in black hole physics. The ringdown waveforms of gravitational waves from two massive objects merging include important information, which is worth studying. The ringdown waveforms are characterized by quasinormal modes (QNMs) of final black holes\cite{Nollert:1999ji, Berti:2007dg, Cardoso:2016rao}. Specially, the ringdown waveforms will no longer depend on the initial disturbance after the initial wave burst of the perturbation. Therefore, the ringdown waveform measurement provide a great chance to reveal properties of the black hole geometry characterized by an event horizon\cite{Price:2017cjr, Giesler:2019uxc}.

	However, QNMs and ringdown signals are in fact related to the photon rings of the massive objects, which may be not necessarily black holes\cite{Cardoso:2016rao}. Thus, it is hard to distinguish black holes from many other horizonless massive objects just via QNMs and ringdown signals\cite{Berti:2009kk, Mark:2017dnq, Konoplya:2011qq}. In fact, many horizonless exotic compact objects (ECOs) can behave like black holes, such as boson stars, gravastars, wormholes and firewalls\cite{Lemos:2008cv, Cunha:2017wao, Cunha:2018acu, Shaikh:2018oul, Dai:2019mse, Huang:2019arj, Simonetti:2020ivl, Wielgus:2020uqz, Yang:2021diz, Bambi:2021qfo, Peng:2021osd, Almheiri:2012rt, Kaplan:2018dqx}. Specially, the late-time ringdown signals together with echoes from the binary black hole merger can be similarly constructed from various ECOs’ theoretical models\cite{Bueno:2017hyj, Konoplya:2018yrp, Wang:2018cum, Wang:2018mlp, Cardoso:2019rvt, GalvezGhersi:2019lag, Liu:2020qia, Ou:2021efv}.
	
	Meanwhile, data analysis suggests that the recent LIGO/Virgo observation may offer potential evidence for echoes in waveforms of gravitational wave from the binary black hole merger\cite{Abedi:2016hgu, Abedi:2017isz}. To gain a deeper insight of echoes, an extra reflecting boundary was laid in a black hole spacetime. It showed that this boundary plays a core role in producing extra multiple time-delay ringdown signals, overlaying to form a series of complete echo waveform which can be received by a distant observer\cite{Mark:2017dnq}. Several black hole and wormhole models have been studied and confirmed to have echoes, such as Kerr-like wormholes, hairy black holes, quantum black holes, braneworld black holes, and even a singularity\cite{Bueno:2017hyj, Bronnikov:2019sbx, Churilova:2019cyt, Liu:2020qia, Ou:2021efv, Konoplya:2018yrp, Kartini:2020ffp, Dey:2020lhq, Pani:2018flj, Urbano:2018nrs, Chowdhury:2020rfj, Wang:2019rcf, Saraswat:2019npa, Oshita:2020dox, Manikandan:2021lko, Liu:2021aqh, DAmico:2019dnn, Guo:2022umh}. Specially, echoes have been found in Einstein -nonlinear electrodynamic spacetimes. One is the dyonic black hole in Einstein-Maxwell gravity extended with quasi-topological electromagnetism, which hold two photon spheres and double-peak effective potentials\cite{Liu:2019rib, Huang:2021qwe}. Another is the multiple-horizon black hole model, which hold effective potentials with three or more peaks\cite{Gao:2021kvr, Sang:2022hng}. Due to the theoretical and observational importance of echoes, it is significant to look for more black hole spacetimes that can hold multiple effective potentials to produce echo signals.
	
	Currently, there is no research about the connection between black hole echoes and quintessence. As recent astronomical observations display, the universe is now in an accelerating expansion\cite{SupernovaCosmologyProject:1998vns, SupernovaSearchTeam:1998fmf, SupernovaSearchTeam:1998cav}, implying a state of negative pressure. Quintessence dark energy is one of the candidates to interpret the negative pressure, whose dynamic may affect the black hole\cite{Kiselev:2002dx, Vagnozzi:2020quf}.
	In this model, the state equation of quintessence is constrained by the pressure $p=\rho_q \omega_q$, where $\rho$ represents the energy density and $\omega_q$ represents the state parameter satisfying $-1<\omega_q<-\frac{1}{3}$. Thus, it is natural for us to think about the effect of quintessence acting to black hole echoes.
	
	The remaining of this paper is organized as follows. In Section \ref{111}, we consider a massless time-dependent scalar field perturbation and offer a numerical method for solving wave function. In Section \ref{222}, we give out the metric of a dyonic black hole with quintessence and analyze its parameters. In Section \ref{333}, we detailly discuss the various transitions of black hole effective potential and echoes under effect of quintessence. In Section \ref{444}, the complete conclusion is given.

	\section{Set up}\label{111}
	\subsection{Massless scalar field perturbation and effective potential}
	
	In this paper, we only consider the static and spherically symmetric black hole. In this section, we give out the most general metric form of the black hole and its effective potential under massless scalar perturbation. The general metric of the black hole takes the form\cite{Huang:2021qwe}
	
	\begin{equation}\label{1}
		ds^2 = -h(r) dt^2 +\frac{1}{f(r)} dr^2 + r^2 (d\theta^2 + \sin ^2 \theta d\phi^2),
	\end{equation}
	where we assume that $h(r_+)=f(r_+)=0$ gives out the event horizon $r_+$ of the black hole. To study the echoes due to the black hole metric only, we consider the time-dependent massless scalar field perturbation, whose equation is
	\begin{equation}\label{2}
		\square \Psi \equiv \frac{1}{\sqrt{-g}} \partial_\mu (\sqrt{-g} g^{\mu\nu} \partial_\nu \Psi(t,r,\theta,\phi)).
	\end{equation}
	
	For the static and spherically symmetric black hole metric, the time-dependent scalar field perturbation $\Psi(t,r,\theta,\phi)$ can be decomposed by separation of variables in terms of spherical harmonics
	\begin{equation}\label{3}
		\Psi(t,r,\theta,\phi) = \sum_{l,m} \frac{\Phi(t,r)}{r} Y_{l,m}(\theta,\phi).
	\end{equation}
	Thus, the scalar field perturbation equation(\ref{1}) reduces to
	\begin{equation}\label{4}
		-\frac{\partial^2 \Phi(t,r)}{\partial t^2} + h(r)f(r) \frac{\partial^2 \Phi(t,r)}{\partial r^2} + \frac{1}{2} \frac{\partial }{\partial r}\left(h(r)f(r)\right) \frac{\partial \Phi(t,r)}{\partial r} - V(r) \Phi(t,r) = 0,
	\end{equation}
	with the original effective potential
	\begin{equation}\label{5}
		V(r) = \frac{l(l+1)}{r^2} h(r) + \frac{1}{2 r} \frac{\partial }{\partial r}\left(h(r)f(r)\right).
	\end{equation}
	In order to map the radial region to $( -\infty, +\infty )$, the definition of the tortoise coordinate $x$ is needed,
	\begin{equation}\label{6}
		dx = \frac{1}{\sqrt{h(r)f(r)}} dr.
	\end{equation}
	Therefore, the equation(\ref{4}) can be rewritten as
	\begin{equation}\label{7}
		\left( -\frac{\partial^2 }{\partial t^2} + \frac{\partial^2 }{\partial x^2}  - V_{eff}(x) \right) \psi(t,x) = 0,
	\end{equation}
	with the time-dependent scalar field perturbation and effective potential in terms of $x$
	\begin{equation}\label{8}
		\Phi(t,r) \Rightarrow \psi(t,x), \quad V(r) \Rightarrow V_{eff}(x).
	\end{equation}
	
	\subsection{Numerical method}
	
	To solve the partial differential equation(\ref{7}), we can consider a time-dependent Green’s function $G(t,x,x')$, which satisfies\cite{Nollert:1999ji, Guo:2022umh}
	\begin{equation}\label{9}
		\left( -\frac{\partial^2 }{\partial t^2} + \frac{\partial^2 }{\partial x^2}  - V_{eff}(x) \right) G(t,x,x') = \delta (t) \delta (x-x').
	\end{equation}
	Thus, the solution of equation(\ref{7}) can be rewritten in terms of $G(t,x,x')$,
	\begin{equation}\label{10}
		\psi(t,x) = - \int_{- \infty}^{+ \infty} \left( G(t,x,x') \partial_t \psi (0,x') + \partial_t G(t,x,x') \psi (0,x') \right) dx'.
	\end{equation}
	With Fourier transform, the solution(\ref{10}) can be expressed as
	\begin{equation}\label{11}
		\psi(t,x) = \frac{1}{2 \pi} \int_{- \infty}^{+ \infty}  \hat{G} (\omega,x,x') \hat{S} (\omega,x) e^{- i \omega t}  dx' d\omega,
	\end{equation}
	where $\hat{S} (\omega,x)$ is depend on the initial condition,
	\begin{equation}
		\hat{S} (\omega,x) =  - \frac{\partial \psi (t,x)}{\partial t} \bigg\vert_{t=0} +i \omega \psi (0,x) ,
	\end{equation}
	
	Two linearly independent solutions $\hat{\psi}_{+}(\omega,x)$ and $\hat{\psi}_{-}(\omega,x)$ are necessary to construct the time-independent Green’s function $G(\omega,x,x')$, with the homogeneous differential equation
	\begin{equation}
		\left( \frac{\partial^2 }{\partial x^2} + \omega^2 - V_{eff}(x) \right) \hat{\psi}(\omega,x) = 0,
	\end{equation}
	and boundary conditions
	\begin{equation}
		\left\{
		\begin{aligned}
			&\hat{\psi}_{+}(\omega,x) = e^{i \omega t}, &&x \rightarrow +\infty \\
			&\hat{\psi}_{-}(\omega,x) = e^{-i \omega t}, \quad &&x \rightarrow -\infty
		\end{aligned}
		\right.
	\end{equation}
	Therefore, the time-independent Green’s function $G(\omega,x,x')$ is given by
	\begin{equation}
		\hat{G}(\omega,x,x') = \frac{ \hat{\psi}_{+} (\omega, \max(x,x'))  \hat{\psi}_{-} (\omega, \min(x,x'))}{W(\omega)} ,
	\end{equation}
	where the Wronskian $W(\omega)$ is defined as
	\begin{equation}
		W(\omega) = \hat{\psi}_{-}(\omega,x) \partial_x \hat{\psi}_{+}(\omega,x) - \partial_x \hat{\psi}_{-}(\omega,x) \hat{\psi}_{+}(\omega,x) ,
	\end{equation}

	\iffalse
	\begin{equation}
		\psi(t,x) = \sum_{n} \psi_n (t,x) = \sum_{n} c_n (x) e^{-i \omega_n t} ,
	\end{equation}
	
	\begin{equation}
		c_n (x) = \frac{-i}{\partial_{\omega} W(\omega_n)} \int_{- \infty}^{+ \infty} \hat{\psi}_{+} (\omega_n, \max(x,x'))  \hat{\psi}_{-} (\omega_n, \min(x,x')) \hat{S}(\omega_n,x') dx' ,
	\end{equation}
	
	\begin{equation}
		\text{Re} \omega_n = \omega_0 + \frac{2 n \pi}{T} ,
	\end{equation}

	\begin{equation}
		\begin{aligned}
			\psi(t,x) &= \sum_{n} \psi_n (t,x) = \sum_{n} c_n (x) e^{-i \omega_n t} \\
			&= 2 \sum_{n} \left( \text{Re} c_n \cos \left( \omega_0 + \frac{2 n \pi}{T} \right) t + \text{Im} c_n \sin \left( \omega_0 + \frac{2 n \pi}{T} \right) t \right) e^{ \text{Im} \omega_n t},
		\end{aligned}
	\end{equation}
	
	\begin{equation}
		\psi_n (t,x) = c_n (x) e^{-i \omega_n t} + c^*_n (x) e^{ - i ( - \omega^*_n ) t} ,
	\end{equation}
	\fi
	
	In this paper, we investigate different waveforms $\psi (t,x)$ caused by different peak structure of effective potential in the black hole, which are all detected by a distant observer. To numerically solve the partial differential equation(\ref{7}) from a certain initial wave packet, we consider the initial condition to be a Gaussian wave packet,
	\begin{equation}
		\psi(0,x)=0, \quad \frac{\partial \psi (t,x)}{\partial t} \bigg\vert_{t=0} = A \exp \left( - \frac{ (x-x_0)^2 }{ 2 \Delta^2 } \right) ,
	\end{equation}
	where the parameter $x_0$ is the center of the initial Gaussian wave packet position, the width $\Delta$ and the amplitude $A$ are also chosen to adapt to the specific case. In this paper, we consider two inequivalent situations. One is that the initial Gaussian wave packet is located outside the double peaks. Another is that it is in the potential well between the two peaks. In \cite{Huang:2021qwe}, it has been found that the black hole echoes have the same characteristics whether it is close to the event horizon or far away from the outer peak. But when the initial location $x_0$ is closer to the bottom of the potential well, the echo frequency roughly doubles the one associated with the wave packet outside the peaks, which makes the echoes more distinguishable. In this paper, we also compare the variation of echo frequency with respect to quintessence parameter in these two cases, which will be discussed in Section \ref{333}.

	\section{Quasi-topological Electromagnetism and dyonic black hole with quintessence}\label{222}
	
	In \cite{Liu:2019rib, Huang:2021qwe}, the quasi-topological electromagnetism is defined to be the squared norm of the topological 4-form $F \land F$. The Lagrangian is
	\begin{equation}
		\mathcal{L} = \sqrt{-g} \left( R-F^{\mu\nu}F_{\mu\nu}-\alpha \left( (F^{\mu\nu}F_{\mu\nu})^2 -2 F^\mu_\nu F^\nu_\rho F^\rho_\sigma F^\sigma_\mu \right) \right).
	\end{equation}
	This theory admits the exact solution of the dyonic black hole, which is both static and spherically symmetric. The Einstein field equation and the Maxwell equation of motion can be expressed as\cite{Liu:2019rib}
	\begin{equation}
		\begin{aligned}
			&R_{\mu \nu} - \frac{1}{2} R g_{\mu \nu} = T_{\mu \nu} + \widetilde{T}_{\mu \nu}, \\
			&\nabla_{\mu} \left( 4 F^{\mu \nu} + 8 \alpha (F^2 F^{\mu \nu} -2 F^{\mu \rho} F^{\sigma}_{rho} F^{\nu}_{\sigma} ) \right) = 0,
		\end{aligned}
	\end{equation}
	where the energy-momentum tensor $T_{\mu \nu}$ can be written as
	\begin{equation}
		T_{\mu \nu} = 2 F_{\mu \rho} F^{\rho}_{\nu} - \frac{1}{2} F^2 g_{\mu \nu} + \alpha \left( 4 F^2 F_{\mu \rho} F^{\rho}_{\nu} - 8 F_{\mu \rho} F^\rho_\sigma F^\sigma_\lambda F^\lambda_\nu - \frac{1}{2} \left( (F^2)^2 -2 F^\mu_\nu F^\nu_\rho F^\rho_\sigma F^\sigma_\mu \right) \right),
	\end{equation}
	and $\widetilde{T}_{\mu \nu}$ denotes the quintessence matter with\cite{Kiselev:2002dx}
	\begin{equation}
		\widetilde{T}^{t}_{t} = \widetilde{T}^{r}_{r} = -\rho_q, \quad \widetilde{T}^{\theta}_{\theta} = \widetilde{T}^{\phi}_{\phi} = - \frac{1}{2} \rho_q (3\omega_q + 1).
	\end{equation}
	Thus, with the above functions, the metric of the black hole with quintessence can be written as
	\begin{equation}
		ds^2 = -f(r) dt^2 +\frac{1}{f(r)} dr^2 + r^2 (d\theta^2 + \sin ^2 \theta d\phi^2),
	\end{equation}
	where
	\begin{equation}\label{25}
		f(r) = 1-\frac{2 M}{r} +\frac{p^2}{r^2} + \frac{q^2}{r^2}  \, _2F_1 \left( \frac{1}{4}, 1; \frac{5}{4}; -\frac{4 \alpha  p^2}{r^4} \right) -\frac{a}{r^{ 3 \omega _q  + 1}}.
	\end{equation}
	With the black hole metric, the original effective potential equation(\ref{5}) reduces to
	\begin{align}
		V(r) &= \frac{l(l+1)}{r^2} f(r) + \frac{1}{r} f(r) \frac{\partial f(r)}{\partial r} \\
		&= \frac{f(r)}{r^2} \left( l(l+1)+\frac{2 M}{r}-\frac{2 p^2}{r^2}-\frac{q^2 r^2}{4 \alpha  p^2+r^4}-\frac{q^2}{r^2}\, _2F_1\left(\frac{1}{4},1;\frac{5}{4};-\frac{4 p^2 \alpha }{r^4}\right) + \frac{a (3 \omega_q+1)}{r^{3 \omega_q+1}} \right). \nonumber
	\end{align}
	Here, we can see that the effective potential $V(r)$ contains several parameters. $M$, $q$ and $p$ are the mass, electric and magnetic charges of the black hole, $\alpha$ is the coupling constant, $\omega_q$ and $a$ are the state parameter and normalization factor of quintessence. When $\alpha=0$ and $p=0$, the metric(\ref{25}) reduces to the usual RN black hole metric. When $q=0$, it reduces to the original Schwarzschild black hole metric.
	
	For a dyonic black hole with quintessence, set $f(r)=0$ and generally we can get three solution. Two of them are black hole horizon, with the larger one becomes the event horizon $r_+$. And the solution far away beyond the event horizon becomes the cosmological horizon $r_c$. The region between the two horizons is called the domain of outer communication, since any two observers in this region may communicate with each other without being hindered by a horizon\cite{Friedman:1993ty, Zeng:2020vsj, He:2021aeo}. Under tortoise coordinates, region $(r_+,r_c)$ turns into $(x_+,x_c)=(-\infty, +\infty)$, which also makes it suitable to consider the cosmological horizon as an observer at infinity.
	
	Moreover, the dyonic black hole has an unusual feature that the metric function $f(r)$ is not monotonous in radius region. It has multiple wiggle such that the gravity force can vanish or even become repulsive in a certain finite region outside the horizon. It also leads to the black hole solution with four or five horizons for some suitable parameters. As this feature has been discussed detailly in \cite{Huang:2021qwe}, we expect that quintessence will give rise to various changes of the black hole horizons and effective potential, which will be analyzed in the following sections.

	\section{Effective potential and black hole echo}\label{333}
	In this section, we discuss different changes of the black hole due to various parameters. After fixing other parameters, we analyze how quintessence influence the number of local maximum and local minimum. It will reduces to the transition of effective potential between single-peak and double-peak, which will have different effect to the echoes of the black hole. Moreover, different values of spherical harmonics $l$ also have different effects on the effective potential. In this paper, we focus on the situation $l = 2$ since they play a dominant role in the ringdown gravitational waves after binary black holes merge.
	
	\subsection{Transition from single-peak to double-peak}
	
	\begin{figure}[htbp]
		\centering
		\includegraphics[scale=0.75]{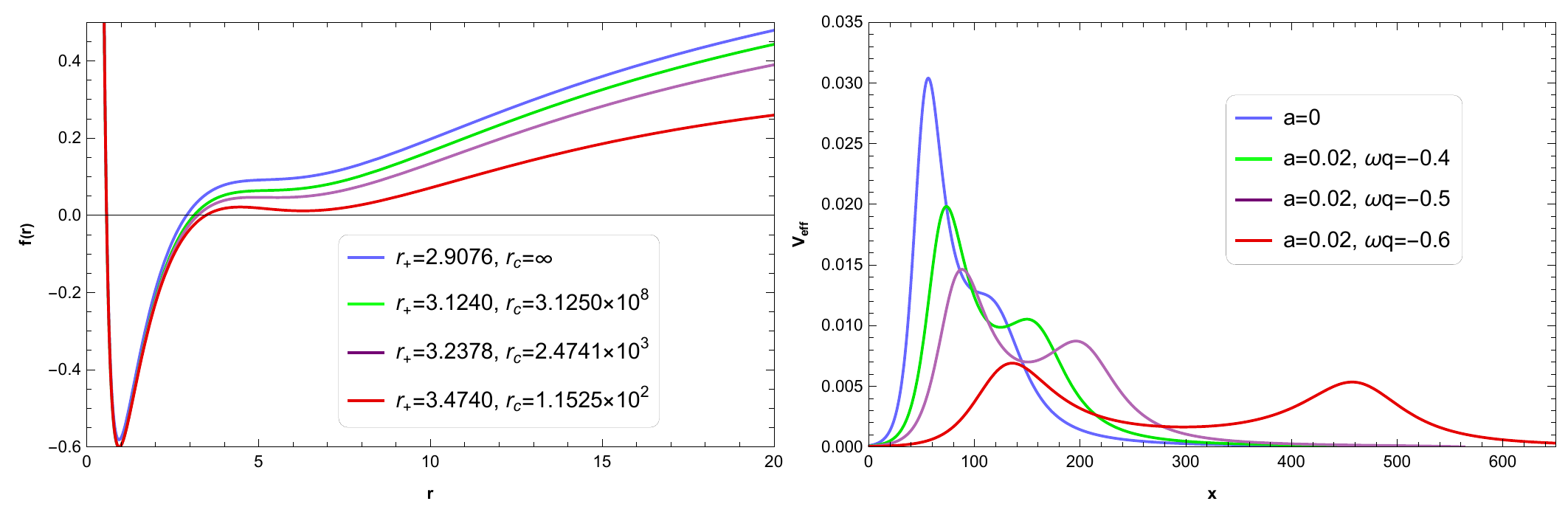}
		\caption{Changes of black hole metric and effective potential under quintessence. Here we take $(M, p, q, \alpha)=(6.44, 1.16, 6.92, 150)$.}
		\label{4.1}
	\end{figure}
	
	\begin{figure}[htbp]
		\centering
		\includegraphics[scale=0.75]{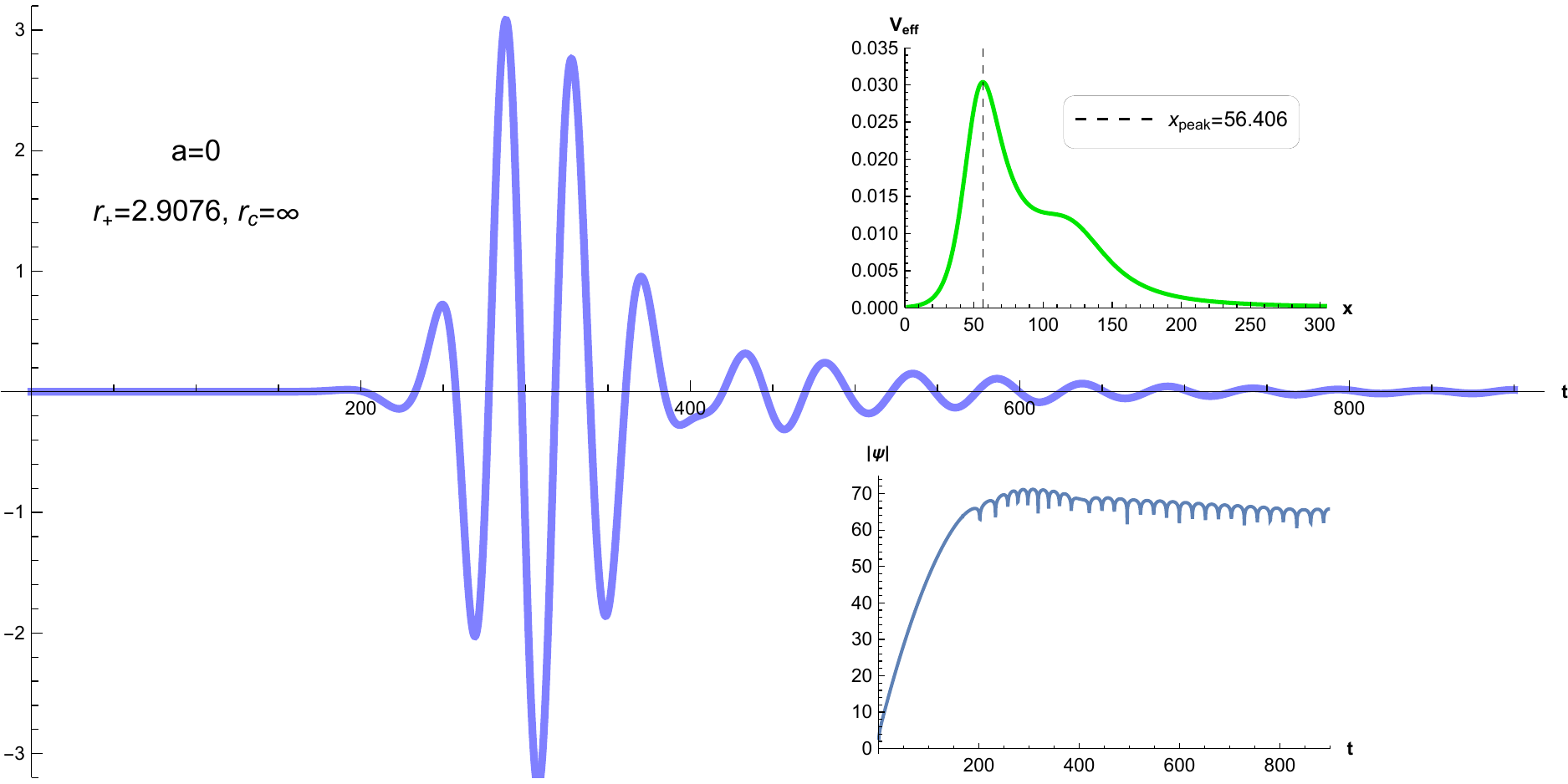}
		\includegraphics[scale=0.75]{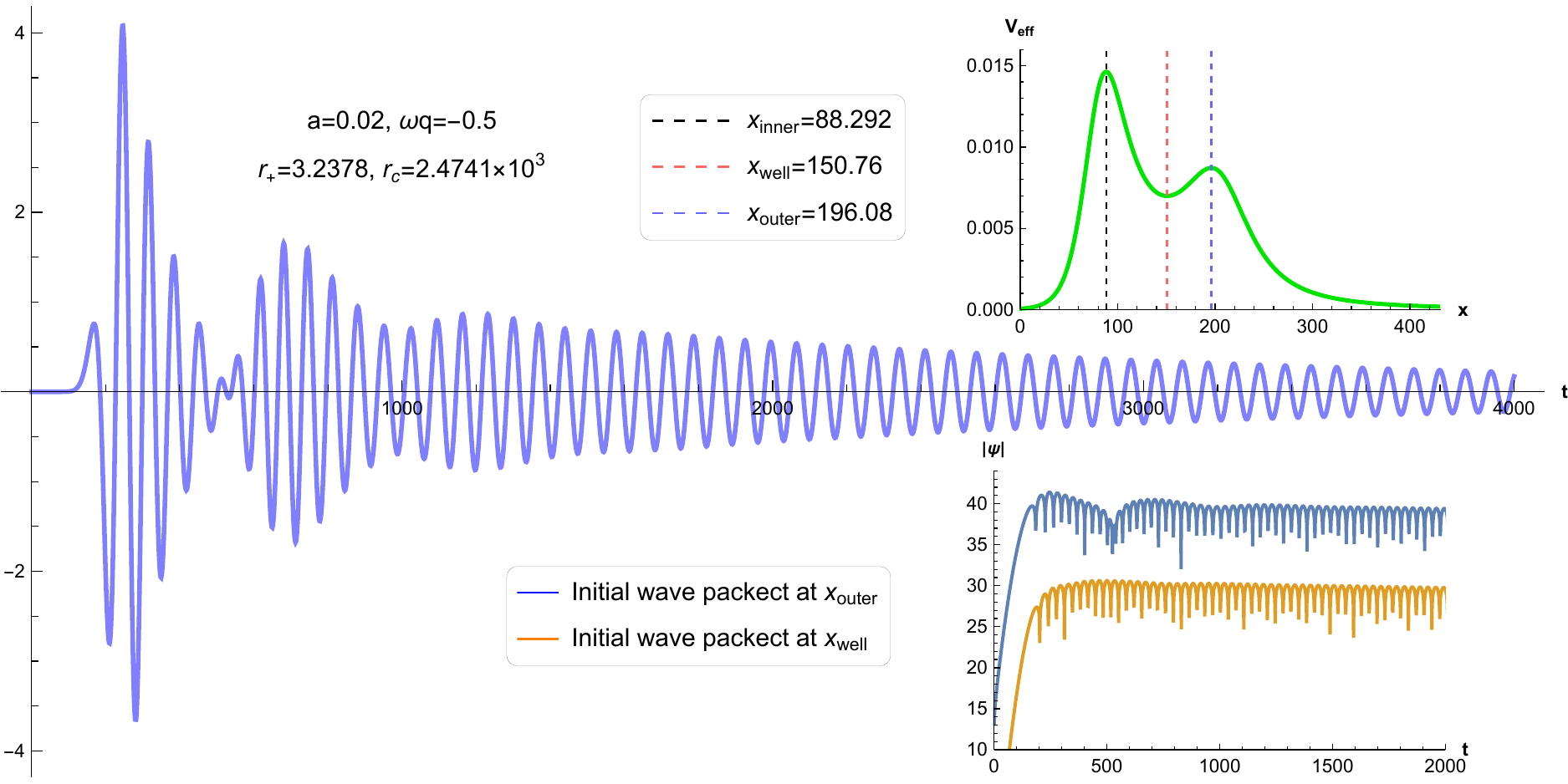}
		\includegraphics[scale=0.75]{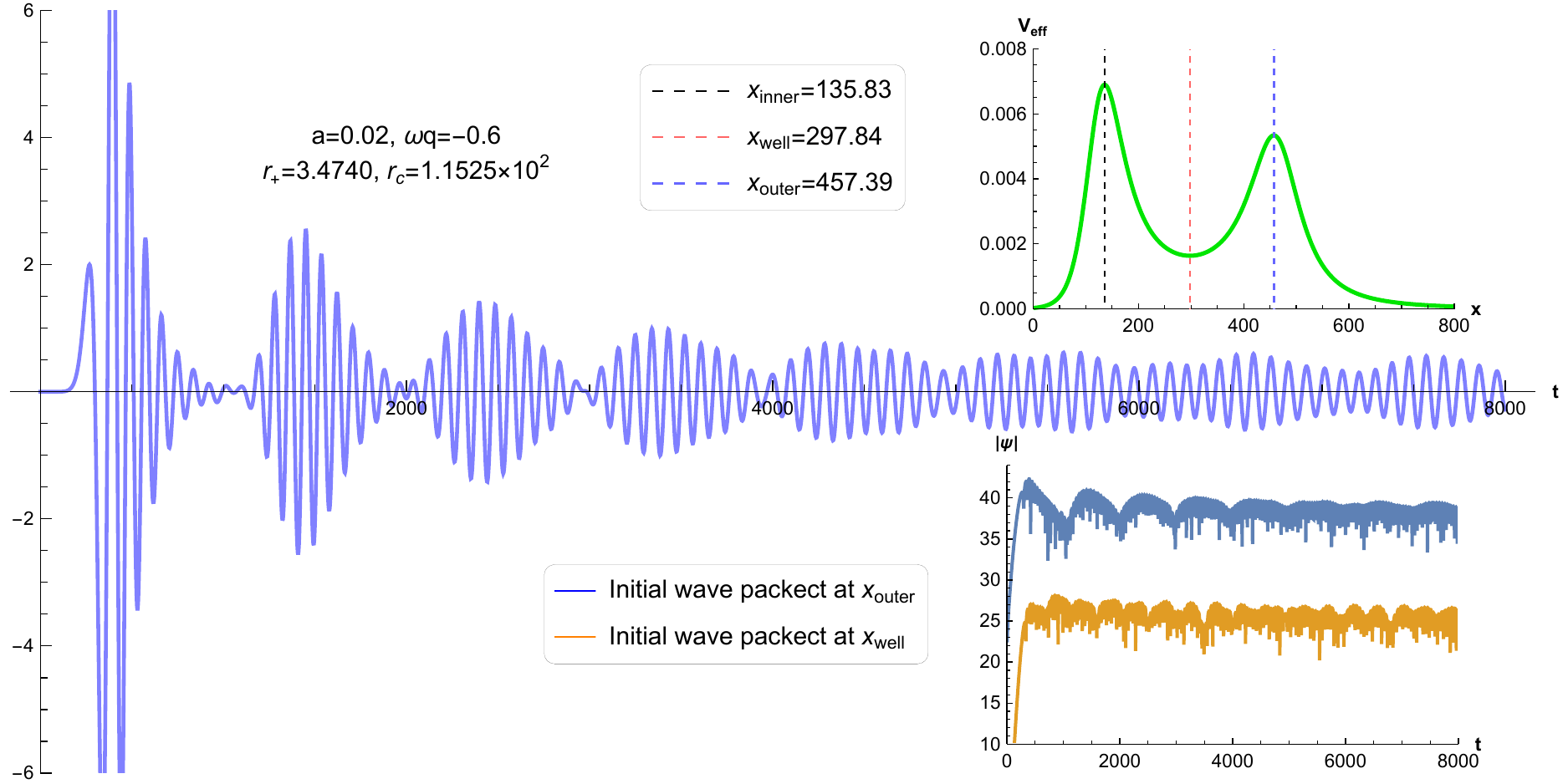}
		\caption{Changes of black hole echoes with $\omega_q$. The upper attached figure gives out the positions and relative size of inner-peak and outer-peak. The lower attached figure gives out echoes from different initial positons of wave package.}
		\label{4.2}
	\end{figure}
	
	For an initial dyonic black hole without quintessence, the coupling constant $\alpha$ is not large enough, meaning that the quasi-topological electromagnetism cannot support strong matter energy-momentum tensor to curve the spacetime. Thus, the effective potential only displays a single-peak, which cannot provide a suitable constructure to generate echoes. The first imiage in Figure \ref{4.2} shows that the reflection from the single-peak potential forms an observed burst after the initial wave traveling from the vicinity of the peak to the observer. At late times, the wave signal shows an exponentially damped sinusoid. Due to the absence of the outer peak, we cannot observe any echo after the burst is received. Note that waves propagating on a black hole spacetime usually develop asymptotically late-time tails, which follow exponentially damped sinusoids and decay as an inverse power of time due to scattering from large radius in the black hole geometry.
	
	After being influenced by quintessence, the effective potential of the black hole reduces continuously and obviously. Meanwhile, as the value of $\omega_q$ becomes small, the outer peak appears and rapidly rises to a size similar to that of the inner peak. Because of the appearance of double-peak, the perturbation can be reflected off by the inner potential barriers and bounce back and forth between the two peaks. When the perturbation successively tunnels through the outer potential barrier, a series of echoes can be received by a distant observer. From Figure \ref{4.2}, we can see that under the effect of quintessence, the relative height of the outer and inner peaks is decreasing, together with the decreasing of the potential well. This means that there will be more wave signals tunneling through the barriers and superposing to form black hole echoes. Thus, the frequency of echoes will increase exponentially under a smaller $\omega_q$, which means a stronger effect from quintessence.

	\subsection{Transition of potential size between inner-peak and outer-peak}
	
	\begin{figure}[htbp]
		\centering
		\includegraphics[scale=0.75]{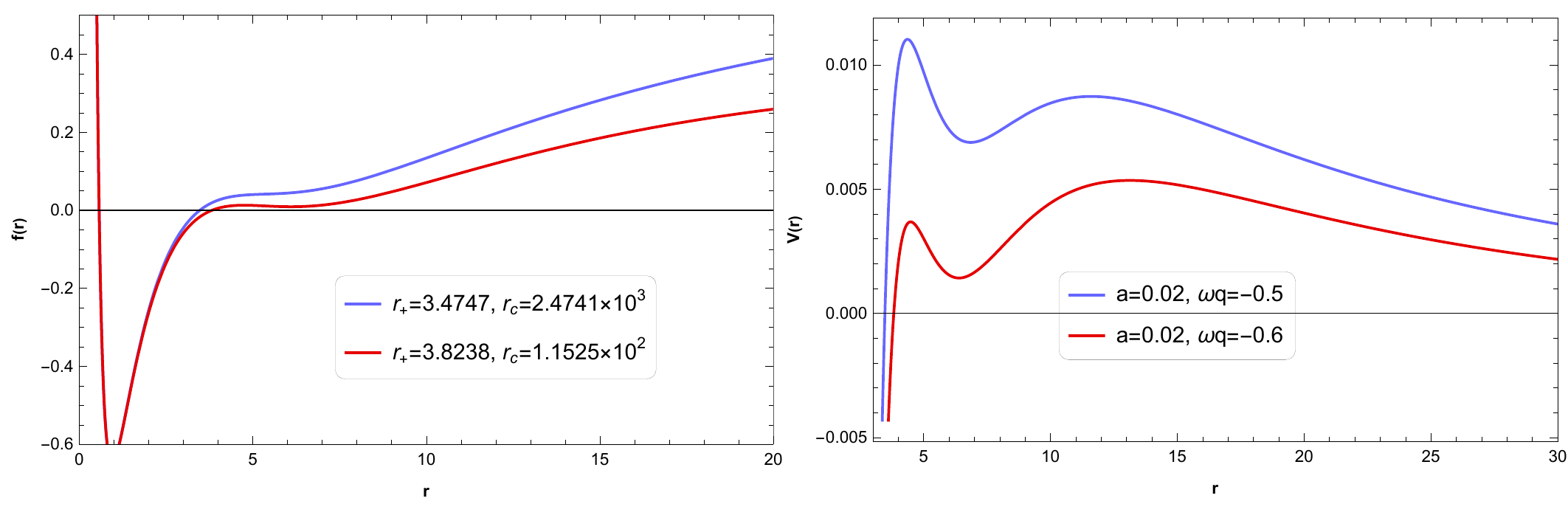}
		\caption{Transition of inner-peak and outer-peak with black hole metric under quintessence. Here we take $(M, p, q, \alpha)=(6.44, 1.185, 6.92, 150)$.}
		\label{4.3}
	\end{figure}
	
	\begin{figure}[htbp]
		\centering
		\includegraphics[scale=0.75]{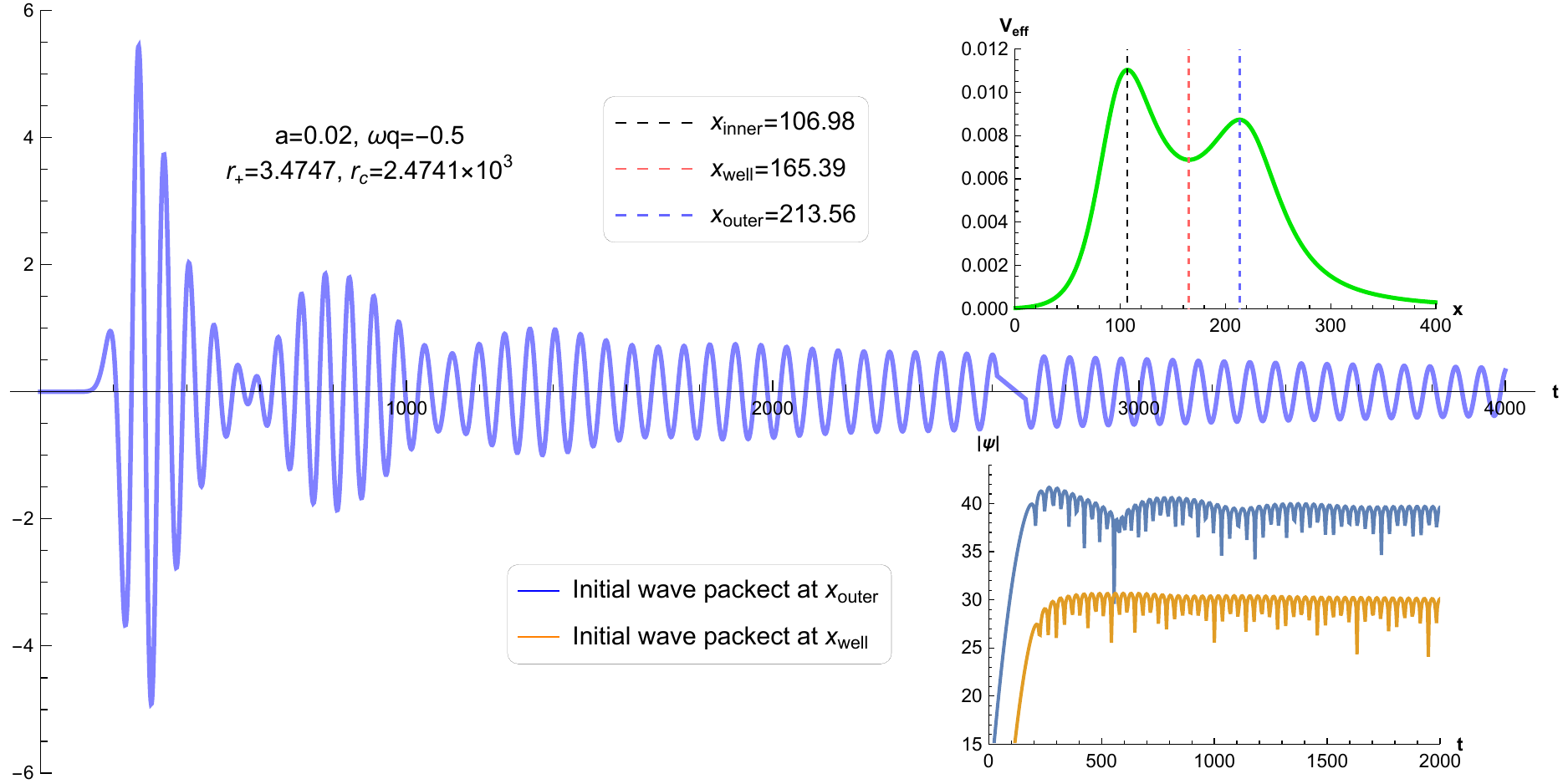}
		\includegraphics[scale=0.75]{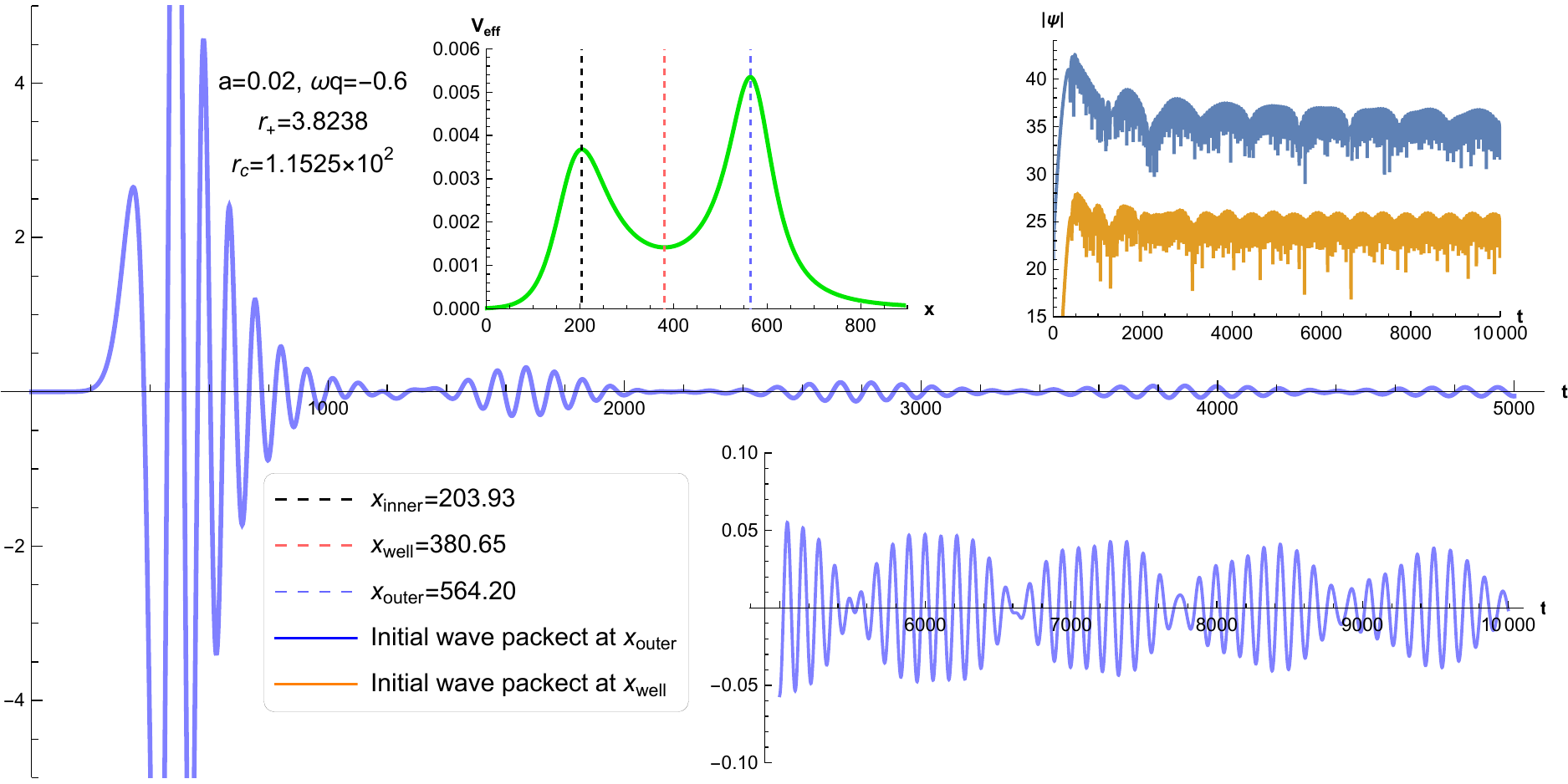}
		\caption{Changes of black hole echoes during the transition of inner-peak and outer-peak with $\omega_q$.}
		\label{4.4}
	\end{figure}

	In the subsection above, we have known that the inner potential peak and potential well will decrease under the effect of quintessence. Therefore, it is natural for us to consider how echoes change when the effective potential changes from an inner-large double-peak to an outer-larger double-peak. When the quasi-topological electromagnetism supports enough matter energy-momentum tensor to curve the spacetime, a dyonic black hole without quintessence can form an effective potential with double-peak. In this case, the distance observer can also receive black hole echoes. And because the inner peak is bigger than the outer peak, the waves reflected by the inner peak can easily tunnel through the outer barrier. This leads to the distinguishable echoes with a high frequency.
	
	However, under the effect of quintessence, the inner peak turns to be lower than the outer one, which influences the behavior of echoes. It should be emphasized that although both the inner potential and potential well are decreasing, the former decreases faster than the latter. As shown in Figure \ref{4.4}, for one thing, as the inner peak decreases, the number of waves that can be reflected by it also decreases accordingly. For another, due to the larger distance between the two peaks, the scattering of a perturbation off one peak is lessly affected by the other peak. Therefore, the black hole echoes become distinguishable but with a low frequency. Meanwhile, waves that has been bounced back need to face a thicker potential barrier, which reduces the probability of tunneling through the barrier. Thus, compared to the initial burst waveform, the amplitude of the echo signal becomes very small, requiring more precise observation instruments.

	\subsection{Smooth and sudden transition from double-peak to single-peak}
	
	\begin{figure}[htbp]
		\centering
		\includegraphics[scale=0.75]{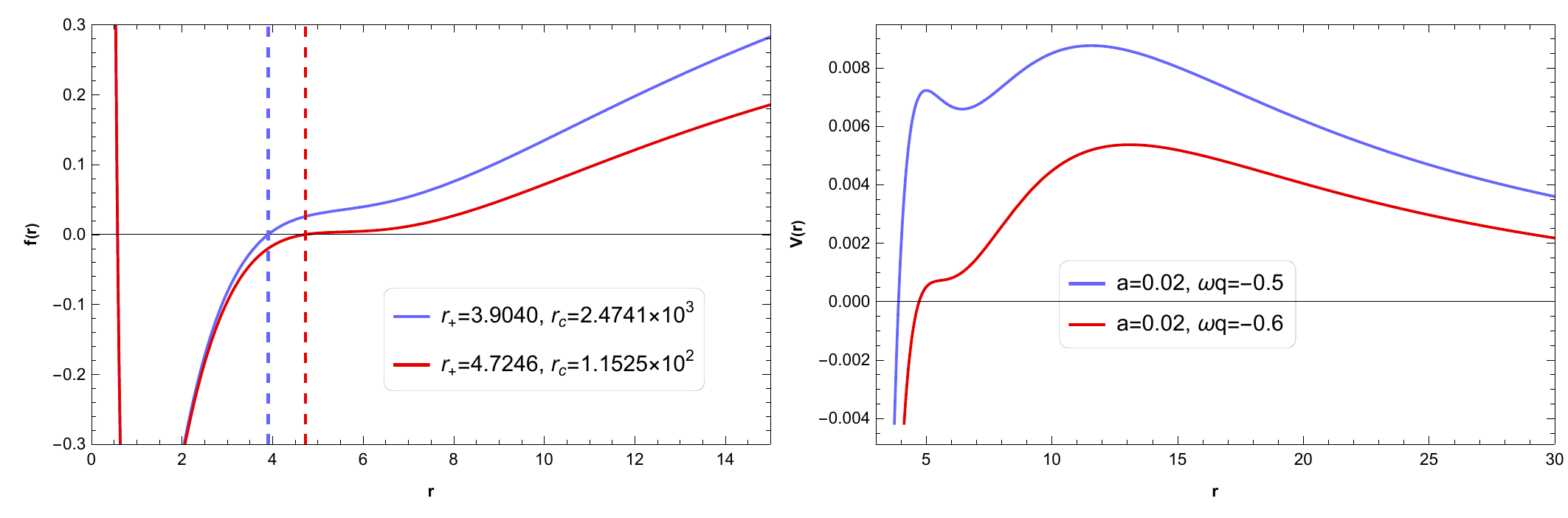}
		\caption{Smooth disappearance of inner-peak with the monotonicity change of black hole metric under quintessence. Here we take $(M, p, q, \alpha)=(6.44, 1.23, 6.92, 150)$.}
		\label{4.5}
	\end{figure}
	
	\begin{figure}[htbp]
		\centering
		\includegraphics[scale=0.75]{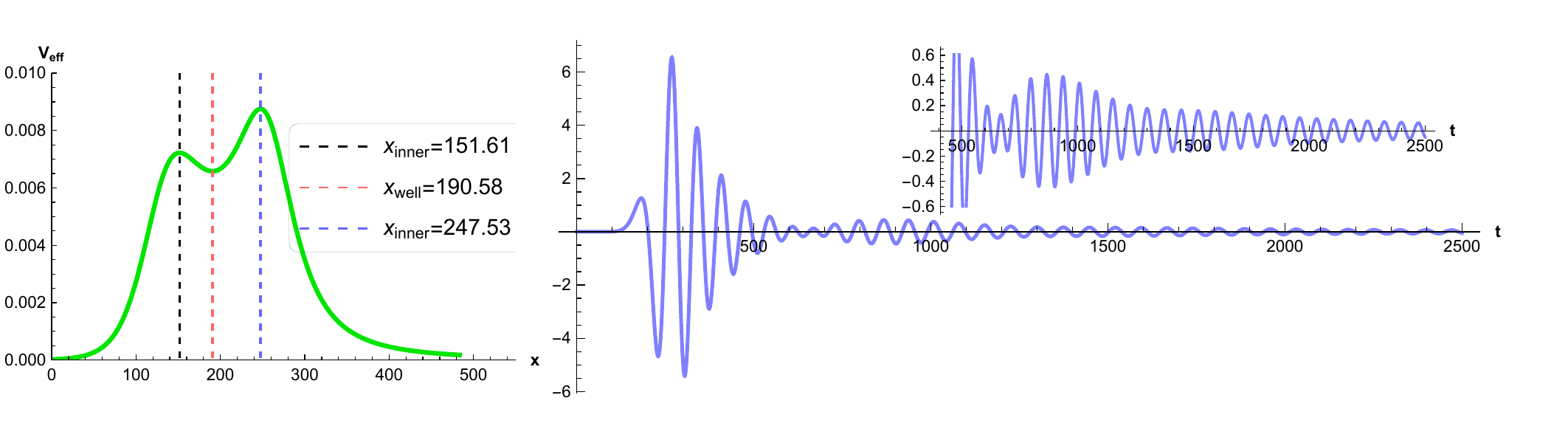}
		\includegraphics[scale=0.75]{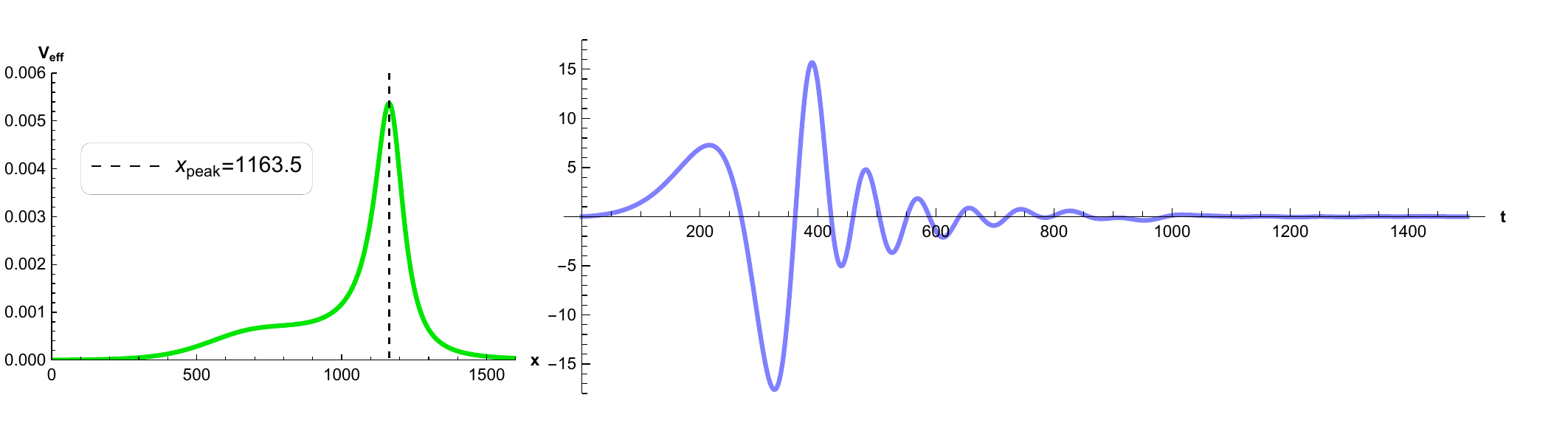}
		\caption{Changes of black hole echoes during the smooth disappearance of inner-peak with $\omega_q$.}
		\label{4.6}
	\end{figure}
	
	\begin{figure}[htbp]
		\centering
		\includegraphics[scale=0.75]{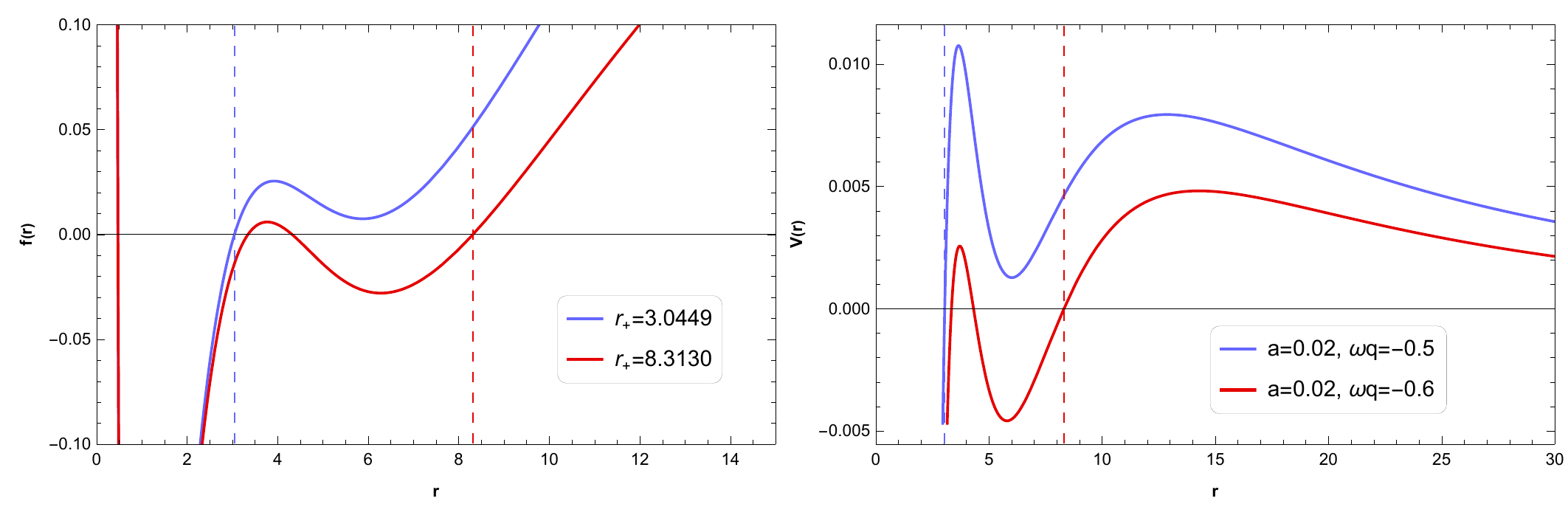}
		\caption{Sudden vanishment of inner-peak with the monotonicity change of black hole metric under quintessence. Here we take $(M, p, q, \alpha)=(6.48, 1.0, 6.8, 150)$.}
		\label{4.7}
	\end{figure}
	
	\begin{figure}[htbp]
		\centering
		\includegraphics[scale=0.75]{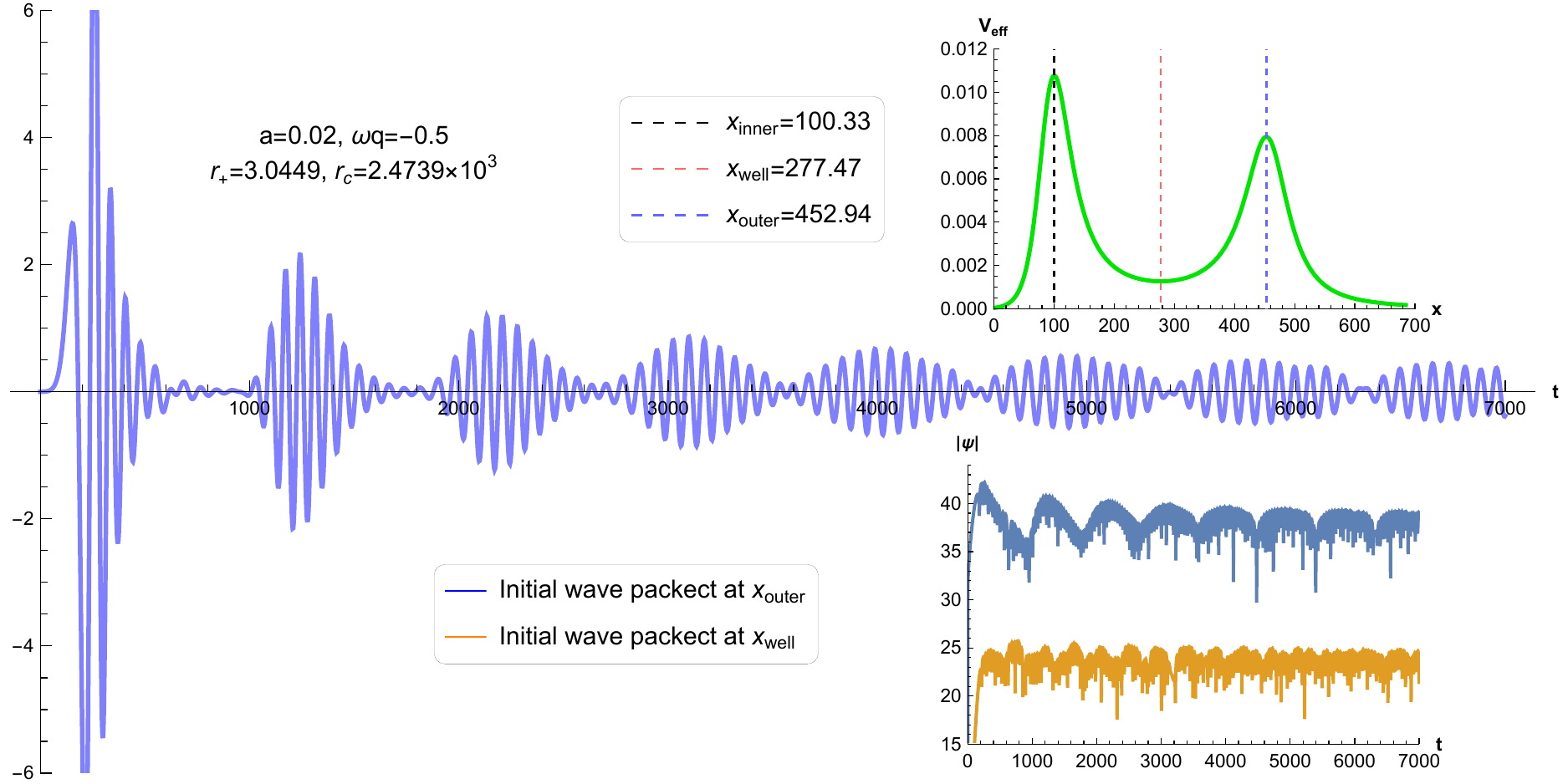}
		\includegraphics[scale=0.75]{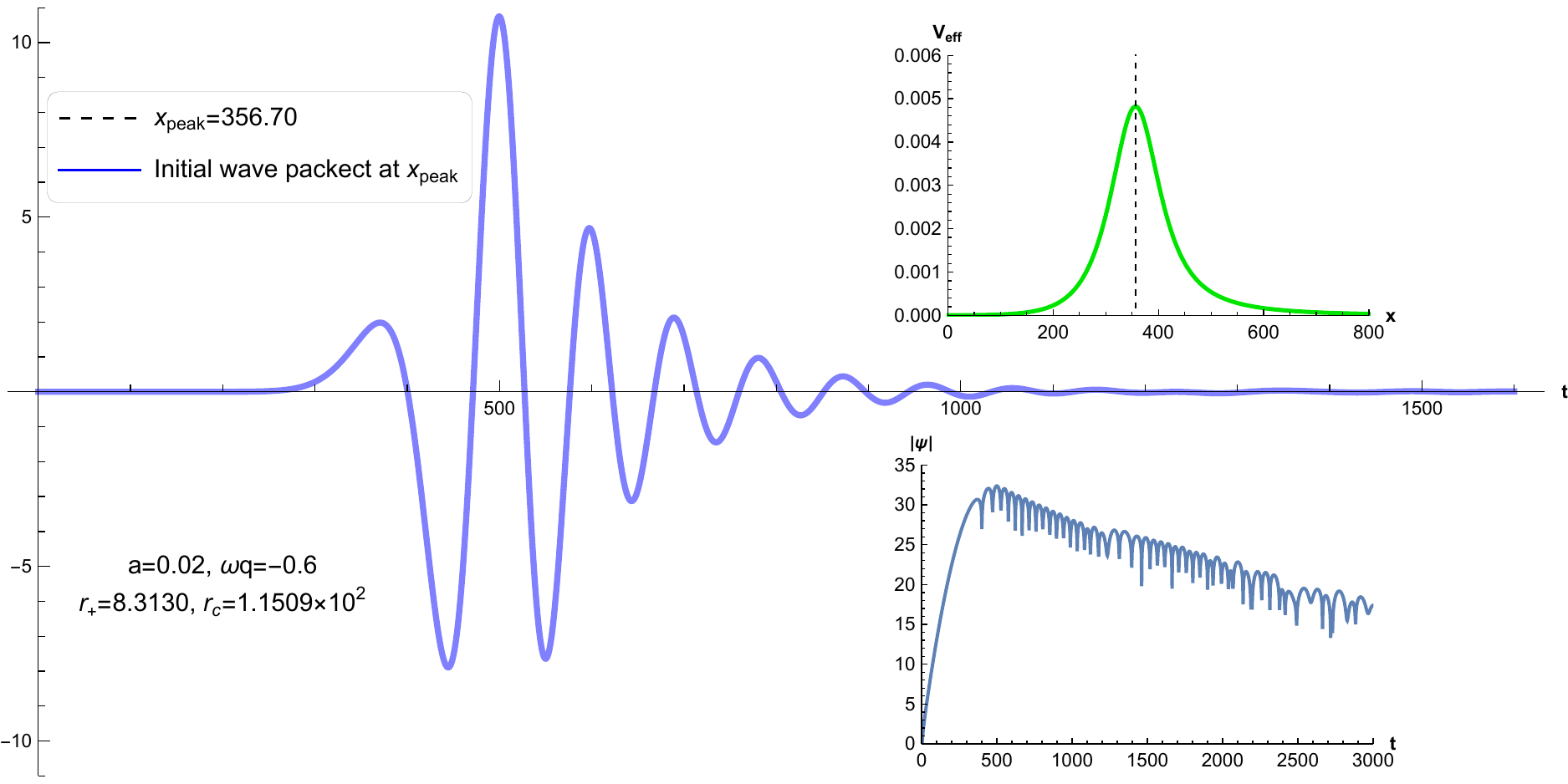}
		\caption{Changes of black hole echoes during the sudden vanishment of inner-peak with $\omega_q$.}
		\label{4.8}
	\end{figure}

	When the effect of quintessence becomes stronger, there will be two situations where the effective potential transforms from double-peak to single-peak. From Figure \ref{4.5}, one situation is that the black hole still maintains three horizons (event horizon $r_+$, cosmology horizon $r_c$ and one horizon inside $r_+$). In this case, because the inner potential decreases faster than the potential well, we can see that the inner peak decreases and finially disappears. Thus, the potential transition is smooth, with the number and frequency of echoes also slowing down gradually from high-frequency and easily resolved echo signal to low-frequency, no-echo waveform signal, which is shown in Figure \ref{4.6}.
	
	From Figure \ref{4.7}, another situation is when the inner potential is large enough, the black hole will have four or five horizons ($r_+$, $r_c$, and two or three horizons inside $r_+$). In this case, the inner peak will be wrapped by event horizon $r_+$ before it disappears. Therefore, as shown in Figure \ref{4.8}, the effective potential will experience a sudden transition from double-peak to single-peak. And the distance observer will also receive a sudden change of the black hole echoes. This change is sudden and rapid. At this moment, distinguishable and high-frequency echoes exist. In the next moment, echoes suddenly disappear, leaving only a single-peak potential.

	\section{Discussion and conclution}\label{444}
	
	In this paper, we first reviewed the dyonic black hole in Einstein-Maxwell gravity extended with quasi-topological electromagnetism which is defined to be the squared norm of the topological 4-form $F \land F$\cite{Liu:2019rib, Huang:2021qwe}. Then we studied various effects of quintessence acting to the black hole effective potential and echoes. Several black hole and wormhole models with double effective potentials have been studied\cite{Bueno:2017hyj, Bronnikov:2019sbx, Churilova:2019cyt, Liu:2020qia, Ou:2021efv, Konoplya:2018yrp, Kartini:2020ffp, Dey:2020lhq, Pani:2018flj, Urbano:2018nrs, Chowdhury:2020rfj, Wang:2019rcf, Saraswat:2019npa, Oshita:2020dox, Manikandan:2021lko, Liu:2021aqh, DAmico:2019dnn, Guo:2022umh}. And our motivation was to look for whether quintessence can influence the spacetime structure of black holes. And our results shown that effective potential indeed varies between single peak and double peaks in a dyonic black hole with quintessence. We also studied echoes with quintessence from different values and distance of inner and outer potential peaks, including wormhole-like potential.
	
%	Moreover, the connection between echoes and spherical harmonics $l$ was discussed, as the quintessence effects are different under different values of $l$.
	
	For suitable parameters, the effect of quintessence acting on the dyonic black hole will cause potential transition between single-peak and double-peak, which can be approximately divided into two cases. One is that an outer peak appears, the original peak becomes the inner one, which decreases and finally generally disappears together with potential well. In this case, the distance observer will receive smooth transition of waveform between exponentially decaying sinusoid and high frequency echoes. Another is that the black hole will exist multiple horizons to wrap the inner peak, which leads to the sudden disappearance of the inner peak. In this case, it will be strange to observe a distinguishable and high-frequency echoes vanish, leaving only an exponentially decaying sinusoid waveform.
	
	It should be emphasized that quintessence only cannot lead to the multiple potential peaks and echoes in Schwarzschild or RN black holes. In fact, the dyonic black hole itself already has double-peak potential under certain conditions. This is caused by Einstein-Maxwell gravity extended with quasi-topological electromagnetism. Under a large coupling constant $\alpha$, the quasi-topological electromagnetism provides proper matter energy-momentum tensor to curve the spacetime\cite{Liu:2019rib, Huang:2021qwe}. And quintessence plays no particular role other than strengthen this force. Thus, it will be interesting if our analysis of quintessence can be extended beyond spherical symmetry to more general black hole spacetimes.

	\section*{Acknowledgement}
	We are grateful to Haitang Yang, Jun Tao and Yuhang Lu for useful discussions. This work is supported in part by NSFC (Grant No. 11747171), Xinglin Scholars Project of Chengdu University of Traditional Chinese Medicine (Grant no.QNXZ2018050).


\begin{thebibliography}{99}
	
	%\cite{LIGOScientific:2016aoc}
	\bibitem{LIGOScientific:2016aoc}
	B.~P.~Abbott \textit{et al.} [LIGO Scientific and Virgo],
	Observation of Gravitational Waves from a Binary Black Hole Merger,
	Phys. Rev. Lett. \textbf{116}, no.6, 061102 (2016)
	doi:10.1103/PhysRevLett.116.061102
	[arXiv:1602.03837 [gr-qc]].
	%9281 citations counted in INSPIRE as of 26 Apr 2023
	
	%\cite{EventHorizonTelescope:2019dse}
	\bibitem{EventHorizonTelescope:2019dse}
	K.~Akiyama \textit{et al.} [Event Horizon Telescope],
	First M87 Event Horizon Telescope Results. I. The Shadow of the Supermassive Black Hole,
	Astrophys. J. Lett. \textbf{875}, L1 (2019)
	doi:10.3847/2041-8213/ab0ec7
	[arXiv:1906.11238 [astro-ph.GA]].
	%2209 citations counted in INSPIRE as of 26 Apr 2023
	
	%\cite{EventHorizonTelescope:2019uob}
	\bibitem{EventHorizonTelescope:2019uob}
	K.~Akiyama \textit{et al.} [Event Horizon Telescope],
	First M87 Event Horizon Telescope Results. II. Array and Instrumentation,
	Astrophys. J. Lett. \textbf{875}, no.1, L2 (2019)
	doi:10.3847/2041-8213/ab0c96
	[arXiv:1906.11239 [astro-ph.IM]].
	%570 citations counted in INSPIRE as of 26 Apr 2023
	
	%\cite{EventHorizonTelescope:2019jan}
	\bibitem{EventHorizonTelescope:2019jan}
	K.~Akiyama \textit{et al.} [Event Horizon Telescope],
	First M87 Event Horizon Telescope Results. III. Data Processing and Calibration,
	Astrophys. J. Lett. \textbf{875}, no.1, L3 (2019)
	doi:10.3847/2041-8213/ab0c57
	[arXiv:1906.11240 [astro-ph.GA]].
	%518 citations counted in INSPIRE as of 26 Apr 2023
	
	%\cite{EventHorizonTelescope:2019ths}
	\bibitem{EventHorizonTelescope:2019ths}
	K.~Akiyama \textit{et al.} [Event Horizon Telescope],
	First M87 Event Horizon Telescope Results. IV. Imaging the Central Supermassive Black Hole,
	Astrophys. J. Lett. \textbf{875}, no.1, L4 (2019)
	doi:10.3847/2041-8213/ab0e85
	[arXiv:1906.11241 [astro-ph.GA]].
	%769 citations counted in INSPIRE as of 26 Apr 2023
	
	%\cite{EventHorizonTelescope:2019pgp}
	\bibitem{EventHorizonTelescope:2019pgp}
	K.~Akiyama \textit{et al.} [Event Horizon Telescope],
	First M87 Event Horizon Telescope Results. V. Physical Origin of the Asymmetric Ring,
	Astrophys. J. Lett. \textbf{875}, no.1, L5 (2019)
	doi:10.3847/2041-8213/ab0f43
	[arXiv:1906.11242 [astro-ph.GA]].
	%808 citations counted in INSPIRE as of 26 Apr 2023
	
	%\cite{EventHorizonTelescope:2019ggy}
	\bibitem{EventHorizonTelescope:2019ggy}
	K.~Akiyama \textit{et al.} [Event Horizon Telescope],
	First M87 Event Horizon Telescope Results. VI. The Shadow and Mass of the Central Black Hole,
	Astrophys. J. Lett. \textbf{875}, no.1, L6 (2019)
	doi:10.3847/2041-8213/ab1141
	[arXiv:1906.11243 [astro-ph.GA]].
	%892 citations counted in INSPIRE as of 26 Apr 2023
	
	%\cite{Nollert:1999ji}
	\bibitem{Nollert:1999ji}
	H.~P.~Nollert,
	TOPICAL REVIEW: Quasinormal modes: the characteristic `sound' of black holes and neutron stars,
	Class. Quant. Grav. \textbf{16}, R159-R216 (1999)
	doi:10.1088/0264-9381/16/12/201
	%753 citations counted in INSPIRE as of 26 Apr 2023
	
	%\cite{Berti:2007dg}
	\bibitem{Berti:2007dg}
	E.~Berti, V.~Cardoso, J.~A.~Gonzalez and U.~Sperhake,
	Mining information from binary black hole mergers: A Comparison of estimation methods for complex exponentials in noise,
	Phys. Rev. D \textbf{75}, 124017 (2007)
	doi:10.1103/PhysRevD.75.124017
	[arXiv:gr-qc/0701086 [gr-qc]].
	%96 citations counted in INSPIRE as of 26 Apr 2023
	
	%\cite{Cardoso:2016rao}
	\bibitem{Cardoso:2016rao}
	V.~Cardoso, E.~Franzin and P.~Pani,
	Is the gravitational-wave ringdown a probe of the event horizon?,
	Phys. Rev. Lett. \textbf{116}, no.17, 171101 (2016)
	[erratum: Phys. Rev. Lett. \textbf{117}, no.8, 089902 (2016)]
	doi:10.1103/PhysRevLett.116.171101
	[arXiv:1602.07309 [gr-qc]].
	%555 citations counted in INSPIRE as of 26 Apr 2023
	
	%\cite{Price:2017cjr}
	\bibitem{Price:2017cjr}
	R.~H.~Price and G.~Khanna,
	Gravitational wave sources: reflections and echoes,
	Class. Quant. Grav. \textbf{34}, no.22, 225005 (2017)
	doi:10.1088/1361-6382/aa8f29
	[arXiv:1702.04833 [gr-qc]].
	%61 citations counted in INSPIRE as of 26 Apr 2023
	
	%\cite{Giesler:2019uxc}
	\bibitem{Giesler:2019uxc}
	M.~Giesler, M.~Isi, M.~A.~Scheel and S.~Teukolsky,
	Black Hole Ringdown: The Importance of Overtones,
	Phys. Rev. X \textbf{9}, no.4, 041060 (2019)
	doi:10.1103/PhysRevX.9.041060
	[arXiv:1903.08284 [gr-qc]].
	%161 citations counted in INSPIRE as of 26 Apr 2023
	
	%\cite{Berti:2009kk}
	\bibitem{Berti:2009kk}
	E.~Berti, V.~Cardoso and A.~O.~Starinets,
	Quasinormal modes of black holes and black branes,
	Class. Quant. Grav. \textbf{26}, 163001 (2009)
	doi:10.1088/0264-9381/26/16/163001
	[arXiv:0905.2975 [gr-qc]].
	%1444 citations counted in INSPIRE as of 26 Apr 2023
	
	%\cite{Mark:2017dnq}
	\bibitem{Mark:2017dnq}
	Z.~Mark, A.~Zimmerman, S.~M.~Du and Y.~Chen,
	A recipe for echoes from exotic compact objects,
	Phys. Rev. D \textbf{96}, no.8, 084002 (2017)
	doi:10.1103/PhysRevD.96.084002
	[arXiv:1706.06155 [gr-qc]].
	%163 citations counted in INSPIRE as of 26 Apr 2023
	
	%\cite{Konoplya:2011qq}
	\bibitem{Konoplya:2011qq}
	R.~A.~Konoplya and A.~Zhidenko,
	Quasinormal modes of black holes: From astrophysics to string theory,
	Rev. Mod. Phys. \textbf{83}, 793-836 (2011)
	doi:10.1103/RevModPhys.83.793
	[arXiv:1102.4014 [gr-qc]].
	%859 citations counted in INSPIRE as of 26 Apr 2023
	
	%\cite{Lemos:2008cv}
	\bibitem{Lemos:2008cv}
	J.~P.~S.~Lemos and O.~B.~Zaslavskii,
	Black hole mimickers: Regular versus singular behavior,
	Phys. Rev. D \textbf{78}, 024040 (2008)
	doi:10.1103/PhysRevD.78.024040
	[arXiv:0806.0845 [gr-qc]].
	%64 citations counted in INSPIRE as of 26 Apr 2023
	
	%\cite{Cunha:2017wao}
	\bibitem{Cunha:2017wao}
	P.~V.~P.~Cunha, J.~A.~Font, C.~Herdeiro, E.~Radu, N.~Sanchis-Gual and M.~Zilh\~ao,
	Lensing and dynamics of ultracompact bosonic stars,
	Phys. Rev. D \textbf{96}, no.10, 104040 (2017)
	doi:10.1103/PhysRevD.96.104040
	[arXiv:1709.06118 [gr-qc]].
	%65 citations counted in INSPIRE as of 26 Apr 2023
	
	%\cite{Cunha:2018acu}
	\bibitem{Cunha:2018acu}
	P.~V.~P.~Cunha and C.~A.~R.~Herdeiro,
	Shadows and strong gravitational lensing: a brief review,
	Gen. Rel. Grav. \textbf{50}, no.4, 42 (2018)
	doi:10.1007/s10714-018-2361-9
	[arXiv:1801.00860 [gr-qc]].
	%301 citations counted in INSPIRE as of 26 Apr 2023
	
	%\cite{Shaikh:2018oul}
	\bibitem{Shaikh:2018oul}
	R.~Shaikh, P.~Banerjee, S.~Paul and T.~Sarkar,
	A novel gravitational lensing feature by wormholes,
	Phys. Lett. B \textbf{789}, 270-275 (2019)
	[erratum: Phys. Lett. B \textbf{791}, 422-423 (2019)]
	doi:10.1016/j.physletb.2018.12.030
	[arXiv:1811.08245 [gr-qc]].
	%62 citations counted in INSPIRE as of 26 Apr 2023
	
	%\cite{Dai:2019mse}
	\bibitem{Dai:2019mse}
	D.~C.~Dai and D.~Stojkovic,
	Observing a Wormhole,
	Phys. Rev. D \textbf{100}, no.8, 083513 (2019)
	doi:10.1103/PhysRevD.100.083513
	[arXiv:1910.00429 [gr-qc]].
	%66 citations counted in INSPIRE as of 26 Apr 2023
	
	%\cite{Huang:2019arj}
	\bibitem{Huang:2019arj}
	H.~Huang and J.~Yang,
	Charged Ellis Wormhole and Black Bounce,
	Phys. Rev. D \textbf{100}, no.12, 124063 (2019)
	doi:10.1103/PhysRevD.100.124063
	[arXiv:1909.04603 [gr-qc]].
	%29 citations counted in INSPIRE as of 26 Apr 2023
	
	%\cite{Simonetti:2020ivl}
	\bibitem{Simonetti:2020ivl}
	J.~H.~Simonetti, M.~J.~Kavic, D.~Minic, D.~Stojkovic and D.~C.~Dai,
	Sensitive searches for wormholes,
	Phys. Rev. D \textbf{104}, no.8, L081502 (2021)
	doi:10.1103/PhysRevD.104.L081502
	[arXiv:2007.12184 [gr-qc]].
	%31 citations counted in INSPIRE as of 26 Apr 2023
	
	%\cite{Wielgus:2020uqz}
	\bibitem{Wielgus:2020uqz}
	M.~Wielgus, J.~Horak, F.~Vincent and M.~Abramowicz,
	Reflection-asymmetric wormholes and their double shadows,
	Phys. Rev. D \textbf{102}, no.8, 084044 (2020)
	doi:10.1103/PhysRevD.102.084044
	[arXiv:2008.10130 [gr-qc]].
	%45 citations counted in INSPIRE as of 26 Apr 2023
	
	%\cite{Yang:2021diz}
	\bibitem{Yang:2021diz}
	J.~Yang and H.~Huang,
	Trapping horizons of the evolving charged wormhole and black bounce,
	Phys. Rev. D \textbf{104}, no.8, 084005 (2021)
	doi:10.1103/PhysRevD.104.084005
	[arXiv:2104.11134 [gr-qc]].
	%8 citations counted in INSPIRE as of 26 Apr 2023
	
	%\cite{Bambi:2021qfo}
	\bibitem{Bambi:2021qfo}
	C.~Bambi and D.~Stojkovic,
	Astrophysical Wormholes,
	Universe \textbf{7}, no.5, 136 (2021)
	doi:10.3390/universe7050136
	[arXiv:2105.00881 [gr-qc]].
	%40 citations counted in INSPIRE as of 26 Apr 2023
	
	%\cite{Peng:2021osd}
	\bibitem{Peng:2021osd}
	J.~Peng, M.~Guo and X.~H.~Feng,
	Observational signature and additional photon rings of an asymmetric thin-shell wormhole,
	Phys. Rev. D \textbf{104}, no.12, 124010 (2021)
	doi:10.1103/PhysRevD.104.124010
	[arXiv:2102.05488 [gr-qc]].
	%38 citations counted in INSPIRE as of 26 Apr 2023
	
	%\cite{Almheiri:2012rt}
	\bibitem{Almheiri:2012rt}
	A.~Almheiri, D.~Marolf, J.~Polchinski and J.~Sully,
	Black Holes: Complementarity or Firewalls?,
	JHEP \textbf{02}, 062 (2013)
	doi:10.1007/JHEP02(2013)062
	[arXiv:1207.3123 [hep-th]].
	%1403 citations counted in INSPIRE as of 26 Apr 2023
	
	%\cite{Kaplan:2018dqx}
	\bibitem{Kaplan:2018dqx}
	D.~E.~Kaplan and S.~Rajendran,
	Firewalls in General Relativity,
	Phys. Rev. D \textbf{99}, no.4, 044033 (2019)
	doi:10.1103/PhysRevD.99.044033
	[arXiv:1812.00536 [hep-th]].
	%24 citations counted in INSPIRE as of 26 Apr 2023
	
	%\cite{Bueno:2017hyj}
	\bibitem{Bueno:2017hyj}
	P.~Bueno, P.~A.~Cano, F.~Goelen, T.~Hertog and B.~Vercnocke,
	Echoes of Kerr-like wormholes,
	Phys. Rev. D \textbf{97}, no.2, 024040 (2018)
	doi:10.1103/PhysRevD.97.024040
	[arXiv:1711.00391 [gr-qc]].
	%152 citations counted in INSPIRE as of 26 Apr 2023
	
	%\cite{Konoplya:2018yrp}
	\bibitem{Konoplya:2018yrp}
	R.~A.~Konoplya, Z.~Stuchl\'\i{}k and A.~Zhidenko,
	Echoes of compact objects: new physics near the surface and matter at a distance,
	Phys. Rev. D \textbf{99}, no.2, 024007 (2019)
	doi:10.1103/PhysRevD.99.024007
	[arXiv:1810.01295 [gr-qc]].
	%58 citations counted in INSPIRE as of 26 Apr 2023
	
	%\cite{Wang:2018cum}
	\bibitem{Wang:2018cum}
	Y.~T.~Wang, J.~Zhang and Y.~S.~Piao,
	Primordial gravastar from inflation,
	Phys. Lett. B \textbf{795}, 314-318 (2019)
	doi:10.1016/j.physletb.2019.06.036
	[arXiv:1810.04885 [gr-qc]].
	%13 citations counted in INSPIRE as of 26 Apr 2023
	
	%\cite{Wang:2018mlp}
	\bibitem{Wang:2018mlp}
	Y.~T.~Wang, Z.~P.~Li, J.~Zhang, S.~Y.~Zhou and Y.~S.~Piao,
	Are gravitational wave ringdown echoes always equal-interval?,
	Eur. Phys. J. C \textbf{78}, no.6, 482 (2018)
	doi:10.1140/epjc/s10052-018-5974-y
	[arXiv:1802.02003 [gr-qc]].
	%30 citations counted in INSPIRE as of 26 Apr 2023
	
	%\cite{Cardoso:2019rvt}
	\bibitem{Cardoso:2019rvt}
	V.~Cardoso and P.~Pani,
	Testing the nature of dark compact objects: a status report,
	Living Rev. Rel. \textbf{22}, no.1, 4 (2019)
	doi:10.1007/s41114-019-0020-4
	[arXiv:1904.05363 [gr-qc]].
	%473 citations counted in INSPIRE as of 26 Apr 2023
	
	%\cite{GalvezGhersi:2019lag}
	\bibitem{GalvezGhersi:2019lag}
	J.~T.~G\'alvez Ghersi, A.~V.~Frolov and D.~A.~Dobre,
	Echoes from the scattering of wavepackets on wormholes,
	Class. Quant. Grav. \textbf{36}, no.13, 135006 (2019)
	doi:10.1088/1361-6382/ab23c8
	[arXiv:1901.06625 [gr-qc]].
	%15 citations counted in INSPIRE as of 26 Apr 2023
	
	%\cite{Liu:2020qia}
	\bibitem{Liu:2020qia}
	H.~Liu, P.~Liu, Y.~Liu, B.~Wang and J.~P.~Wu,
	Echoes from phantom wormholes,
	Phys. Rev. D \textbf{103}, no.2, 024006 (2021)
	doi:10.1103/PhysRevD.103.024006
	[arXiv:2007.09078 [gr-qc]].
	%24 citations counted in INSPIRE as of 26 Apr 2023
	
	%\cite{Ou:2021efv}
	\bibitem{Ou:2021efv}
	M.~Y.~Ou, M.~Y.~Lai and H.~Huang,
	Echoes from asymmetric wormholes and black bounce,
	Eur. Phys. J. C \textbf{82}, no.5, 452 (2022)
	doi:10.1140/epjc/s10052-022-10421-x
	[arXiv:2111.13890 [gr-qc]].
	%13 citations counted in INSPIRE as of 26 Apr 2023
	
	%\cite{Abedi:2016hgu}
	\bibitem{Abedi:2016hgu}
	J.~Abedi, H.~Dykaar and N.~Afshordi,
	Echoes from the Abyss: Tentative evidence for Planck-scale structure at black hole horizons,
	Phys. Rev. D \textbf{96}, no.8, 082004 (2017)
	doi:10.1103/PhysRevD.96.082004
	[arXiv:1612.00266 [gr-qc]].
	%287 citations counted in INSPIRE as of 26 Apr 2023
	
	%\cite{Abedi:2017isz}
	\bibitem{Abedi:2017isz}
	J.~Abedi, H.~Dykaar and N.~Afshordi,
	Echoes from the Abyss: The Holiday Edition!,
	[arXiv:1701.03485 [gr-qc]].
	%91 citations counted in INSPIRE as of 26 Apr 2023
	
	%\cite{Bronnikov:2019sbx}
	\bibitem{Bronnikov:2019sbx}
	K.~A.~Bronnikov and R.~A.~Konoplya,
	Echoes in brane worlds: ringing at a black hole--wormhole transition,
	Phys. Rev. D \textbf{101}, no.6, 064004 (2020)
	doi:10.1103/PhysRevD.101.064004
	[arXiv:1912.05315 [gr-qc]].
	%46 citations counted in INSPIRE as of 26 Apr 2023
	
	%\cite{Churilova:2019cyt}
	\bibitem{Churilova:2019cyt}
	M.~S.~Churilova and Z.~Stuchlik,
	Ringing of the regular black-hole/wormhole transition,
	Class. Quant. Grav. \textbf{37}, no.7, 075014 (2020)
	doi:10.1088/1361-6382/ab7717
	[arXiv:1911.11823 [gr-qc]].
	%49 citations counted in INSPIRE as of 26 Apr 2023
	
	%\cite{Kartini:2020ffp}
	\bibitem{Kartini:2020ffp}
	D.~Kartini and A.~Sulaksono,
	Gravitational wave echoes from quark stars,
	J. Phys. Conf. Ser. \textbf{1572}, 012034 (2020)
	doi:10.1088/1742-6596/1572/1/012034
	%2 citations counted in INSPIRE as of 26 Apr 2023
	
	%\cite{Dey:2020lhq}
	\bibitem{Dey:2020lhq}
	R.~Dey, S.~Chakraborty and N.~Afshordi,
	Echoes from braneworld black holes,
	Phys. Rev. D \textbf{101}, no.10, 104014 (2020)
	doi:10.1103/PhysRevD.101.104014
	[arXiv:2001.01301 [gr-qc]].
	%43 citations counted in INSPIRE as of 26 Apr 2023
	
	%\cite{Pani:2018flj}
	\bibitem{Pani:2018flj}
	P.~Pani and V.~Ferrari,
	On gravitational-wave echoes from neutron-star binary coalescences,
	Class. Quant. Grav. \textbf{35}, no.15, 15LT01 (2018)
	doi:10.1088/1361-6382/aacb8f
	[arXiv:1804.01444 [gr-qc]].
	%42 citations counted in INSPIRE as of 26 Apr 2023
	
	%\cite{Urbano:2018nrs}
	\bibitem{Urbano:2018nrs}
	A.~Urbano and H.~Veerm\"ae,
	On gravitational echoes from ultracompact exotic stars,
	JCAP \textbf{04}, 011 (2019)
	doi:10.1088/1475-7516/2019/04/011
	[arXiv:1810.07137 [gr-qc]].
	%46 citations counted in INSPIRE as of 26 Apr 2023
	
	%\cite{Chowdhury:2020rfj}
	\bibitem{Chowdhury:2020rfj}
	A.~Chowdhury and N.~Banerjee,
	Echoes from a singularity,
	Phys. Rev. D \textbf{102}, no.12, 124051 (2020)
	doi:10.1103/PhysRevD.102.124051
	[arXiv:2006.16522 [gr-qc]].
	%9 citations counted in INSPIRE as of 26 Apr 2023
	
	%\cite{Wang:2019rcf}
	\bibitem{Wang:2019rcf}
	Q.~Wang, N.~Oshita and N.~Afshordi,
	Echoes from Quantum Black Holes,
	Phys. Rev. D \textbf{101}, no.2, 024031 (2020)
	doi:10.1103/PhysRevD.101.024031
	[arXiv:1905.00446 [gr-qc]].
	%64 citations counted in INSPIRE as of 26 Apr 2023
	
	%\cite{Saraswat:2019npa}
	\bibitem{Saraswat:2019npa}
	K.~Saraswat and N.~Afshordi,
	Quantum Nature of Black Holes: Fast Scrambling versus Echoes,
	JHEP \textbf{04}, 136 (2020)
	doi:10.1007/JHEP04(2020)136
	[arXiv:1906.02653 [hep-th]].
	%19 citations counted in INSPIRE as of 26 Apr 2023
	
	%\cite{Oshita:2020dox}
	\bibitem{Oshita:2020dox}
	N.~Oshita, D.~Tsuna and N.~Afshordi,
	Quantum Black Hole Seismology I: Echoes, Ergospheres, and Spectra,
	Phys. Rev. D \textbf{102}, no.2, 024045 (2020)
	doi:10.1103/PhysRevD.102.024045
	[arXiv:2001.11642 [gr-qc]].
	%20 citations counted in INSPIRE as of 26 Apr 2023
	
	%\cite{Manikandan:2021lko}
	\bibitem{Manikandan:2021lko}
	S.~K.~Manikandan and K.~Rajeev,
	New kind of echo from quantum black holes,
	Phys. Rev. D \textbf{105}, no.6, 064024 (2022)
	doi:10.1103/PhysRevD.105.064024
	[arXiv:2112.08773 [gr-qc]].
	%3 citations counted in INSPIRE as of 26 Apr 2023
	
	%\cite{Liu:2021aqh}
	\bibitem{Liu:2021aqh}
	H.~Liu, W.~L.~Qian, Y.~Liu, J.~P.~Wu, B.~Wang and R.~H.~Yue,
	Alternative mechanism for black hole echoes,
	Phys. Rev. D \textbf{104}, no.4, 044012 (2021)
	doi:10.1103/PhysRevD.104.044012
	[arXiv:2104.11912 [gr-qc]].
	%18 citations counted in INSPIRE as of 26 Apr 2023
	
	%\cite{DAmico:2019dnn}
	\bibitem{DAmico:2019dnn}
	G.~D'Amico and N.~Kaloper,
	Black hole echoes,
	Phys. Rev. D \textbf{102}, no.4, 044001 (2020)
	doi:10.1103/PhysRevD.102.044001
	[arXiv:1912.05584 [gr-qc]].
	%10 citations counted in INSPIRE as of 26 Apr 2023
	
	%\cite{Guo:2022umh}
	\bibitem{Guo:2022umh}
	G.~Guo, P.~Wang, H.~Wu and H.~Yang,
	Echoes from hairy black holes,
	JHEP \textbf{06}, 073 (2022)
	doi:10.1007/JHEP06(2022)073
	[arXiv:2204.00982 [gr-qc]].
	%7 citations counted in INSPIRE as of 26 Apr 2023
	
	%\cite{Liu:2019rib}
	\bibitem{Liu:2019rib}
	H.~S.~Liu, Z.~F.~Mai, Y.~Z.~Li and H.~L\"u,
	Quasi-topological Electromagnetism: Dark Energy, Dyonic Black Holes, Stable Photon Spheres and Hidden Electromagnetic Duality,
	Sci. China Phys. Mech. Astron. \textbf{63}, 240411 (2020)
	doi:10.1007/s11433-019-1446-1
	[arXiv:1907.10876 [hep-th]].
	%44 citations counted in INSPIRE as of 26 Apr 2023
	
	%\cite{Huang:2021qwe}
	\bibitem{Huang:2021qwe}
	H.~Huang, M.~Y.~Ou, M.~Y.~Lai and H.~Lu,
	Echoes from classical black holes,
	Phys. Rev. D \textbf{105}, no.10, 104049 (2022)
	doi:10.1103/PhysRevD.105.104049
	[arXiv:2112.14780 [hep-th]].
	%7 citations counted in INSPIRE as of 26 Apr 2023
	
	%\cite{Gao:2021kvr}
	\bibitem{Gao:2021kvr}
	C.~Gao,
	Black holes with many horizons in the theories of nonlinear electrodynamics,
	Phys. Rev. D \textbf{104}, no.6, 064038 (2021)
	doi:10.1103/PhysRevD.104.064038
	[arXiv:2106.13486 [gr-qc]].
	%8 citations counted in INSPIRE as of 26 Apr 2023
	
	%\cite{Sang:2022hng}
	\bibitem{Sang:2022hng}
	A.~Sang, M.~Zhang, S.~W.~Wei and J.~Jiang,
	Echoes of black holes in Einstein-nonlinear electrodynamic theories,
	Eur. Phys. J. C \textbf{83}, no.4, 291 (2023)
	doi:10.1140/epjc/s10052-023-11448-4
	[arXiv:2209.02559 [gr-qc]].
	%0 citations counted in INSPIRE as of 26 Apr 2023
	
	%\cite{SupernovaCosmologyProject:1998vns}
	\bibitem{SupernovaCosmologyProject:1998vns}
	S.~Perlmutter \textit{et al.} [Supernova Cosmology Project],
	Measurements of $\Omega$ and $\Lambda$ from 42 high redshift supernovae,
	Astrophys. J. \textbf{517}, 565-586 (1999)
	doi:10.1086/307221
	[arXiv:astro-ph/9812133 [astro-ph]].
	%14996 citations counted in INSPIRE as of 26 Apr 2023
	
	%\cite{SupernovaSearchTeam:1998fmf}
	\bibitem{SupernovaSearchTeam:1998fmf}
	A.~G.~Riess \textit{et al.} [Supernova Search Team],
	Observational evidence from supernovae for an accelerating universe and a cosmological constant,
	Astron. J. \textbf{116}, 1009-1038 (1998)
	doi:10.1086/300499
	[arXiv:astro-ph/9805201 [astro-ph]].
	%15162 citations counted in INSPIRE as of 26 Apr 2023
	
	%\cite{SupernovaSearchTeam:1998cav}
	\bibitem{SupernovaSearchTeam:1998cav}
	P.~M.~Garnavich \textit{et al.} [Supernova Search Team],
	Supernova limits on the cosmic equation of state,
	Astrophys. J. \textbf{509}, 74-79 (1998)
	doi:10.1086/306495
	[arXiv:astro-ph/9806396 [astro-ph]].
	%782 citations counted in INSPIRE as of 26 Apr 2023
	
	%\cite{Kiselev:2002dx}
	\bibitem{Kiselev:2002dx}
	V.~V.~Kiselev,
	Quintessence and black holes,
	Class. Quant. Grav. \textbf{20}, 1187-1198 (2003)
	doi:10.1088/0264-9381/20/6/310
	[arXiv:gr-qc/0210040 [gr-qc]].
	%471 citations counted in INSPIRE as of 26 Apr 2023
	
	%\cite{Vagnozzi:2020quf}
	\bibitem{Vagnozzi:2020quf}
	S.~Vagnozzi, C.~Bambi and L.~Visinelli,
	Concerns regarding the use of black hole shadows as standard rulers,
	Class. Quant. Grav. \textbf{37}, no.8, 087001 (2020)
	doi:10.1088/1361-6382/ab7965
	[arXiv:2001.02986 [gr-qc]].
	%69 citations counted in INSPIRE as of 26 Apr 2023
	
	%\cite{Friedman:1993ty}
	\bibitem{Friedman:1993ty}
	J.~L.~Friedman, K.~Schleich and D.~M.~Witt,
	Topological censorship,
	Phys. Rev. Lett. \textbf{71}, 1486-1489 (1993)
	[erratum: Phys. Rev. Lett. \textbf{75}, 1872 (1995)]
	doi:10.1103/PhysRevLett.71.1486
	[arXiv:gr-qc/9305017 [gr-qc]].
	%426 citations counted in INSPIRE as of 18 May 2023
	
	%\cite{Zeng:2020vsj}
	\bibitem{Zeng:2020vsj}
	X.~X.~Zeng and H.~Q.~Zhang,
	Influence of quintessence dark energy on the shadow of black hole,
	Eur. Phys. J. C \textbf{80}, no.11, 1058 (2020)
	doi:10.1140/epjc/s10052-020-08656-7
	[arXiv:2007.06333 [gr-qc]].
	%75 citations counted in INSPIRE as of 18 May 2023
	
	%\cite{He:2021aeo}
	\bibitem{He:2021aeo}
	A.~He, J.~Tao, Y.~Xue and L.~Zhang,
	Shadow and photon sphere of black hole in clouds of strings and quintessence *,
	Chin. Phys. C \textbf{46}, no.6, 065102 (2022)
	doi:10.1088/1674-1137/ac56cf
	[arXiv:2109.13807 [gr-qc]].
	%17 citations counted in INSPIRE as of 18 May 2023
	
	\end{thebibliography}
\end{document}